\documentclass{aastex62}
\usepackage{natbib}
\usepackage{amsmath}
\usepackage{graphicx}

\newcommand\gp{$g^{\prime}$}
\newcommand\rp{$r^{\prime}$}
\newcommand\ip{$i^{\prime}$}

\received{January 1, 2018}
\revised{January 7, 2018}
\accepted{\today}
\submitjournal{AJ}

\shorttitle{OGLE-2018-BLG-0022}
\shortauthors{ROME/REA Team}
\begin{document}

\title{OGLE-2018-BLG-0022: A Nearby M-dwarf Binary}

\correspondingauthor{R.A.~Street}
\email{rstreet@lco.global}

\author{R.A.~Street}
\affil{LCOGT, 6740 Cortona Drive, Sutie 102, Goleta, CA 93117, USA.}

\author{E.~Bachelet}
\affiliation{LCOGT, 6740 Cortona Drive, Sutie 102, Goleta, CA 93117, USA.}

\author{Y.~Tsapras}
\affiliation{Zentrum f{\"u}r Astronomie der Universit{\"a}t Heidelberg, Astronomisches Rechen-Institut, M{\"o}nchhofstr. 12-14, 69120 Heidelberg, Germany}

\author{M.P.G.~Hundertmark}
\affiliation{Zentrum f{\"u}r Astronomie der Universit{\"a}t Heidelberg, Astronomisches Rechen-Institut, M{\"o}nchhofstr. 12-14, 69120 Heidelberg, Germany}

\author{V.~Bozza}
\affiliation{Dipartimento di Fisica "E.R. Canianiello", Universit{\'a} di Salerno, Via Giovanni Paolo II 132, 84084, Fisciano, Italy}

\author{M.~Dominik}
\affiliation{Centre for Exoplanet Science, SUPA, School of Physics \& Astronomy, University of St Andrews, North Haugh, St Andrews KY16 9SS, UK}
\collaboration{ROME/REA and MiNDSTEp Teams}

\collaboration{ROME/REA Team}
\author{D.M.~Bramich}
\affiliation{New York University Abu Dhabi, Saadiyat Island, Abu Dhabi, PO Box 129188, United Arab Emirates}

\author{A.~Cassan}
\affiliation{Institut d’Astrophysique de Paris, Sorbonne Universit\'e, CNRS, UMR 7095, 98 bis bd Arago, 75014 Paris, France}

\author{K. Horne}
\affiliation{Centre for Exoplanet Science, SUPA, School of Physics \& Astronomy, University of St Andrews, North Haugh, St Andrews KY16 9SS, UK}

\author{S.~Mao}
\affiliation{Physics Department and Tsinghua Centre for Astrophysics, Tsinghua University, Beijing 100084, China}
\affiliation{National Astronomical Observatories, Chinese Academy of Sciences, 20A Datun Road, Chaoyang District, Beijing 100012, China}

\author{A.~Saha}
\affiliation{National Optical Astronomy Observatory, 950 North Cherry Ave., Tucson, AZ 85719, USA}

\author{J.~Wambsganss}
\affiliation{Zentrum f{\"u}r Astronomie der Universit{\"a}t Heidelberg, Astronomisches Rechen-Institut, M{\"o}nchhofstr. 12-14, 69120 Heidelberg, Germany}
\affiliation{International Space Science Institute (ISSI), Hallerstra{\ss}e 6, 3012 Bern, Switzerland}

\author[0000-0001-6000-3463]{Weicheng~Zang}
\affiliation{Physics Department and Tsinghua Centre for Astrophysics, Tsinghua University, Beijing 100084, China}

\collaboration{MiNDSTEp Team}
\author{U.G.~J\/{o}rgensen}
\affiliation{Niels Bohr Institute \& Centre for Star and Planet Formation, University of Copenhagen, \/{O}ster Voldgade 5, 1350 Copenhagen, Denmark}

\author{P.~Longa-Pe\~{n}a}
\affiliation{Unidad de Astronom\'{i}a, Universidad de Antofagasta, Av. Angamos 601, Antofagasta, Chile}

\author{N.~Peixinho}
\affiliation{CITEUC - Center for Earth and Space Research of the University of Coimbra, Geophysical and Astronomical Observatory, R. Observatorio s/n, 3040-004 Coimbra, Portugal}

\author{S.~Sajadian}
\affiliation{Department~of~Physics,~Isfahan~University~of~Technology,~Isfahan~84156-83111,~Iran}

\author{M.J.~Burgdorf}
\affiliation{Universit\"{a}t Hamburg, Faculty of Mathematics, Informatics and Natural Sciences, Department of Earth Sciences, Meteorological Institute, Bundesstra{\ss}e 55, 20146 Hamburg, Germany}

\author{J.~Campbell-White}
\affiliation{Centre for Astrophysics \& Planetary Science, The University of Kent, Canterbury CT2 7NH, UK}

\author{S.~Dib}
\affiliation{Max Planck Institute for Astronomy, K\"{o}nigstuhl 17, D-69117, Heidelberg, Germany}

\author{D.F.~Evans}
\affiliation{Astrophysics Group, Keele University, Staffordshire, ST5 5BG, UK}

\author{Y.I.~Fujii}
\affiliation{Department of Physics, Nagoya University, Furo-cho, Chikusa-ku, Nagoya, 464-8602, Japan}
\affiliation{Institute for Advanced Research, Nagoya University, Furo-cho, Chikusa-ku, Nagoya, 464-8601, Japan}

\author{T.C.~Hinse}
\affiliation{Chungnam National University, Department of Astronomy and Space Science, 34134 Daejeon, Republic of Korea}

\author{E.~Khalouei}
\affiliation{Department of Physics, Sharif University of Technology, PO Box 11155-9161 Tehran, Iran}

\author{S.~Lowry}
\affiliation{Centre for Astrophysics \& Planetary Science, The University of Kent, Canterbury CT2 7NH, UK}

\author{S.~Rahvar}
\affiliation{Department of Physics, Sharif University of Technology, PO Box 11155-9161 Tehran, Iran}

\author{M.~Rabus}
\affiliation{Instituto de Astronom\'\i{}a, Facultad de F\'\i{}sica, Pontificia Universidad Cat\'olica de Chile, casilla 306, Santiago 22, Chile, Max-Planck-Institut f\"ur Astronomie, K\"onigstuhl 17, D-69117 Heidelberg, Germany, LCOGT, 6740 Cortona Dr., Suite 102, Goleta, CA 93111, USA, Department of Physics, University of California, Santa Barbara, CA 93106-9530, USA}

\author{J.~Skottfelt}
\affiliation{Centre for Electronic Imaging, Department of Physical Sciences, The Open University, Milton Keynes, MK7 6AA, UK}

\author{C.~Snodgrass}
\affiliation{School of Physical Sciences, Faculty of Science, Technology, Engineering and Mathematics, The Open University, Walton Hall, Milton Keynes, MK7 6AA, UK}
\affiliation{Institute for Astronomy, University of Edinburgh, Royal Observatory, Edinburgh EH9 3HJ, UK}

\author{J.~Southworth}
\affiliation{Astrophysics Group, Keele University, Staffordshire, ST5 5BG, UK}

\author{J.~Tregloan-Reed}
\affiliation{Unidad de Astronom\'{i}a, Universidad de Antofagasta, Av. Angamos 601, Antofagasta, Chile}



\begin{abstract}

We report observations of the binary microlensing event OGLE-2018-BLG-0022, provided by the ROME/REA Survey, which indicate that the lens is a low-mass binary star consisting of M3 (0.375$\pm$0.020\,M$_{\odot}$) and M7 (0.098$\pm$0.005\,M$_{\odot}$) components.  The lens is unusually close, at 0.998$\pm$0.047\,kpc, compared with the majority of microlensing events, and despite its intrinsically low luminosity, it is likely that AO observations in the near future will be able to provide an independent confirmation of the lens masses.  

\end{abstract}



\section{Introduction} 
\label{sec:intro}
Microlensing offers a way to explore the populations of stellar and planetary systems in regions of the Galaxy where they are too faint to study via alternative techniques, and at orbital separations where reflex-based and transit methods are inefficient.  Seventy-two planetary systems discovered by their lensing signature have been published to date\footnote{Source: NASA Exoplanet Archive, https://exoplanetarchive.ipac.caltech.edu/} but notably the method is also sensitive to many other intrinsically low-luminosity objects, including late-type stars and brown dwarfs as well as compact objects, including white dwarfs and black holes \citep{Wyrzykowski2016}, since the technique depends on the gravity, rather than the light, from the lensing system. 

A microlensing event occurs when a foreground object crosses the observer's line of sight to an unrelated luminous source in the background, causing the latter to brighten and fade as the objects move into and out of alignment.  Since the events are transient, occurring unpredictably\footnote{We note that astrometry from the Gaia Mission has recently enabled some events to be predicted in advance \citep{Bramich2018} but still for a relatively limited sample of stars.} and without repetition, surveys typically maximize their yield by photometric monitoring of densely populated regions of the Galactic Bulge, where the microlensing optical depth, or probability of lensing is greatest,
$\Gamma$=[18.74$\pm$0.91]$\times$10$^{-6}$exp[(0.53$\pm$0.05)(3-$|b|$)] star$^{-1}$ yr$^{-1}$ for $|l|<5^{\circ}$ \citep{SumiPenny2016}, resulting in $\sim$2000 events being discovered per year, of which $\sim$10\,\% are due to binary lenses.  While the guiding scientific goal of most of these surveys is generally the discovery of exoplanets, they yield binary lens systems with a wide range of mass ratios, all of which must be carefully observed and assessed to determine the true nature of the lensing system.  

Here we present multi-band observations of the microlensing event OGLE-2018-BLG-0022 from the new ROME/REA survey, along with a description of the analysis process.  In the next section, we outline the essential theoretical model parameters and our motivation for this observing strategy, followed by a brief description of the ROME/REA project and observations of this event.  The light curve modeling and analyses are presented in Sections~\ref{sec:modeling} and \ref{sec:colour_analysis} and we discuss their implications for the nature of the lens in Section~\ref{sec:lens}.  

\section{Characterizing Microlensing Events}
\label{sec:theory}
A foreground lensing object of mass $M_{L}$, at distance $D_{L}$ from the observer, deflects the light from a background source at distance $D_{S}$ with a characteristic angular radius, $\theta_{E} = \sqrt{\frac{4GM_{L}}{c^{2}}\frac{D_{LS}}{D_{L}D_{S}}}$ \citep{Refsdal1964}, where $D_{LS}$ is the distance between the lens and source.  As the relative proper motion, $\mu_{\rm{rel}}$ of the lens and source narrows their projected angular separation $u(t)$ to a minimum $u_{0}$ at time $t_{0}$, the source appears magnified as a function of time, with the magnification given by $A(t) = \frac{u^{2} + 2}{u\sqrt{u^{2} + 4}}$.  Microlensing events have a characteristic Einstein crossing time, $t_{E}$, defined as the time taken for the source to cross $\theta_{E}$ in a lens-centered geometry. 

At their simplest, single, point lens microlensing events are described by just three parameters,  $t_{0}$, $u_{0}$, $t_{E}$, and binary lenses require just three more: the mass ratio of the lens components, $q = M_{L,2}/M_{L,1}$, their angular separation, $s$, normalized by $\theta_{E}$ and $\alpha$, the counterclockwise angle between the binary axis and the source trajectory. 

All of these parameters may be measured directly from time-series photometry in a single passband, but unfortunately this alone does not reveal the physical nature of the lens, since $\theta_{E}$ has a mass-distance degeneracy \citep{Dominik1999}.  This ambiguity is most commonly broken by measuring two effects.  

The motion of the observer during the event requires a modification of $u(t)$ to take microlensing parallax, $\pi_{E} = (\pi_{E,N}, \pi_{E,E})$, into account.  This may be measured as a skew in the light curve of events with $t_{E} \gtrsim 30$\,d, or otherwise by combining simultaneous light curves from widely separated observers, such as on Earth and in space (e.g. \citealt{Dong2007, Shvartzvald2016}).  Although both lens and source may be kiloparsecs distant from the observer, the finite angular size of the latter can nevertheless introduce detectable distortions around the peak of the light curve, parameterized as $\rho = \theta_{S}/\theta_{E}$.  $\rho$ can then be used to determine $\theta_{E}$, if an independent measurement is made of the angular radius of the source $\theta_{S}$. 

As microlensing sources are typically faint, with $I \gtrsim 15.0$\,mag, their angular sizes are most easily estimated from stellar models based on their spectral type.  This is usually constrained from a low-cadence light curve of the event in a second optical bandpass, since the microlensing magnification can be used to distinguish the light from the source from other stars blended within the same Point Spread Function (PSF).   Ongoing microlensing surveys, such as the Optical Gravitational Lensing Experiment (OGLE\footnote{http://ogle.astrouw.edu.pl/} \citealt{Udalski1992ogleproject}),  Microlensing Observations in Astrophysics  (MOA\footnote{http://www.phys.canterbury.ac.nz/moa/, \citealt{Sako2008moaproject, Bond2001, Sumi2003})} and Korea Microlensing Telescope Network (KMTNet, \citealt{Park2012KMTNetproject}), typically obtain imaging data in two broadband filters, usually Bessell-$V, I$.  Priority is given to $I$-band observations in order to properly constrain all light curve features, with $V$-band data obtained at a much lower and variable cadence.  

\section{The ROME/REA project}
\label{sec:romerea}
The goal of the ROME/REA Microlensing Project (described in \citet{Tsapras2018romerea}) is to ensure that the source stars of microlensing events within its footprint are well characterized and hence that the physical nature of the lensing objects can be determined.  The project has adopted a novel observing strategy designed to complement those of the existing surveys, which combines both regular survey-mode observations (ROME, Robotic Observations of Microlensing Events) in three passbands with higher cadence single-filter (REA, REActive mode) observations obtained around the event peaks, or in response to caustic crossings.  This strategy takes advantage of the multiple 1\,m telescopes at each site of the Las Cumbres Observatory Telescope Network (LCO) and the flexibility offered by the network's robotic scheduling system \citep{Saunders2014}.  

The ROME survey monitors 20 selected fields in the Galactic Bulge where the rate of microlensing events is highest \citep{SumiPenny2016}.  The field of view of each pointing is $26'\times26'$, determined by the field of the {\it Sinistro} cameras of the LCO 1\,m network, giving a total survey footprint of $3.76$\, sq. deg. A triplet of 300\,s exposures in SDSS-g$^{\prime}$, -r$^{\prime}$ and -i$^{\prime}$ are obtained in each survey visit to a field, and all 20 fields are surveyed with a nominal cadence of once every 7\,hrs thanks to the geographic distribution of the LCO network \citep{Brown2013LCOGT}.  Specifically, ROME/REA uses the LCO Southern Ring of identical 1\,m telescopes at Cerro Tololo Inter-American Observatory (CTIO), Chile, the South African Astronomical Observatory (SAAO), South Africa and Siding Spring Observatory (SSO), Australia.  ROME survey observations are therefore conducted around the clock, as long as the fields are visible from each site, between April 1 to October 31 each year, starting in 2017.  

As such, the ROME survey was designed to complement other ongoing surveys, by improving the color data available to characterize microlensing source stars and filling a gap between the surveys that observe the Bulge at high cadence but predominantly in a single filter and very wide-field surveys that obtain multi-bandpass data but sometimes at a cadence that is too low to provide useful constraints to microlensing events.  For example, OGLE and KMTNet obtain data in $V$ at $<$1\,d cadence but $I$ band data at intervals $<$15\,min, while the Zwicky Transient Factory observes the northern Plane nightly in SDSS-r and occasionally in SDSS-g.  ROME/REA complements the wavelength coverage of the NIR UKIRT  \citep{Shvartzvald2017} and VVV surveys \citep{Minniti2010}. 

\section{Observations and Data Reduction} 
\label{sec:obs}

The event OGLE-2018-BLG-0022 was first discovered and classified as a microlensing event by OGLE on 2018 February 7, and subsequently re-identified by the same survey as OGLE-2018-BLG-0052 on 2018 February 21.  The same object was also independently discovered by MOA on 2018 February 25, who assigned the label MOA-2018-BLG-031.  

With RA, Dec coordinates of 17:59:27.04, -28:36:37.00 (J2000.0), this event lies within the boundaries of ROME-FIELD-16.  ROME observations of this field began on 2017 March 18 using the LCO facilities summarized in Table~\ref{tab:telescopes}.  In general, we endeavored to conduct ROME and REA observations using a consistent set of cameras at the 3 sites in order to limit the number of datasets and any calibration offsets between them, so the majority of our data was provided by 3 instruments.  However, the LCO network is designed to optimize its schedule globally by moving observation requests between telescopes, and REA-mode observations in particular were obtained from multiple cameras for this reason.  Over the longer term, it was also necessary occasionally to transfer ROME observations between telescopes at the same site, when technical issues affected the original instruments.  Nevertheless, all data were obtained using the {\it Sinistro} class of optical cameras, all of which consist of 4k$\times$4k Fairchild CCDs operated in bin 1$\times$1 mode with a pixel scale of 0.389\arcsec pix$^{-1}$.  

\begin{table}[h!]
\centering
\caption{Summary of telescopes and instruments used} \label{tab:telescopes}
\begin{tabular}{lllll}
\hline
\hline
Obs. Mode & Site         & Telescope & Camera    & Filters \\
\hline
ROME      & Chile 		 & Dome C, 1m0-04 & fl03 & \ip \\
ROME      & Chile 		 & Dome A, 1m0-05 & fl15 & \gp , \rp, \ip \\
ROME      & South Africa & Dome A, 1m0-10 & fl16 & \gp , \rp, \ip \\
ROME	  & South Africa & Dome C, 1m0-12 & fl06 & \rp, \ip \\
ROME	  & Australia	 & Dome A, 1m0-11 & fl12 & \gp , \rp, \ip \\
REA       & Chile 		 & Dome A, 1m0-05 & fl15 & \gp, \rp, \ip\\
REA       & Chile 		 & Dome C, 1m0-04 & fl03 & \ip \\
REA       & South Africa & Dome A, 1m0-10 & fl16 & \ip \\
REA		  & South Africa & Dome C, 1m0-12 & fl06 & \rp, \ip \\
REA		  & Australia    & Dome A, 1m0-11 & fl12 & \gp, \rp, \ip \\
REA		  & Australia    & Dome B, 1m0-03 & fl11 & \ip \\
MiNDSTEp  & Chile        & 1.54\,m        & EMCCD & $i’_{\rm DK}$ \\
\hline
\multicolumn{4}{l}{Total number of images}            & 1260 \\
\hline
\end{tabular}
\end{table}

On 2018 March 13 the ARTEMiS anomaly detection system \citep{Dominik2008ARTEMiS} found that the light curve of the event was deviating from a point-source, point-lens model on the rising section of its light curve, and subsequent modeling efforts by Bozza, Cassan, Bachelet and Hirao\footnote{Private communications} confirmed that the event was most likely caused by a binary lens.  As the event brightened towards its peak magnification it met the criteria for REA and our RoboTAP target prioritization software \citep{Hundertmark2018RoboTAP} began to schedule REA-mode observations in addition to those for ROME. The models provided by V. Bozza's RTModel system for real-time analysis \citep{Bozza2010} provided predictions regarding the timing of future caustic crossings that were used to plan observations.  Following the ROME/REA strategy, REA-LO mode, single-filter observations were automatically requested every hour, while REA-HI observations were triggered to ensure data would be obtained at high-cadence (every 15\,min) for the periods of predicted caustic crossing. Photometry was provided to RTModel from several teams including ROME/REA while the event was in progress, which allowed both the model predictions and the REA observations to be updated accordingly until the event was observed to return to the source's baseline brightness. REA-mode observations continued until after the peak of the event, ending on 2018 June 10.

All ROME/REA imaging data were preprocessed by the standard LCO {\em BANZAI} pipeline to remove the instrumental signatures, then reduced using a Difference Image Analysis (DIA) pipeline based on the {\em DanDIA} package by \cite{Bramich2008, Bramich2013} to produce light curve photometry. 

Independently of ROME/REA, MiNDSTEp observations with the Danish 1.54,m in Chile were triggered automatically by the SIGNALMEN anomaly detector \citep{Dominik2007}, operated as part of the ARTEMiS system\footnote{http://www.artemis-uk.org/} \citep{Dominik2008,Dominik2010}, in conjunction with real-time modeling of anomalous events provided by RTModel\footnote{http://www.fisica.unisa.it/GravitationAstrophysics/RTModel.htm} \citep{Bozza2018}.  They began on 2018 April 25 and continued until 2018 May 18, with the goal of ensuring high-cadence coverage of the anomaly.  These data were obtained with the EMCCD camera equipped with a long-pass filter with a short-wavelength cut off at 6500\,\AA, making the filter function resemble a combined SDSS-$i^{\prime}$ plus SDSS-$z^{\prime}$ plus the long-wavelength part of the SDSS-$r^{\prime}$ filter, denoted as $i_{DK}$ in Table~\ref{tab:telescopes}. These data were reduced with a version of the {\em DanDIA} package \citep{Bramich2008} which has been optimized for the reduction of data from this EMCCD instrument \citep{Skottfelt2015, Evans2016}.  

\section{light curve Analysis}
\label{sec:modeling}

Some residual structures remained after the initial processing. As the event timescale is relatively long ($t_{E} > 50$\,d), it was likely that annual parallax and, potentially, the orbital motion of the lens may be significant.  We therefore explore these two second-order effects and find a great improvement of the model likelihood.

Since the light curve presents clear signatures of a multiple lens, we began by fitting a simple Uniform-Source Binary Lens (USBL) model to the light curve data, where both lens and observer were considered to be static, using the pyLIMA modeling package \citep{Bachelet2017}.  It should be noted that pyLIMA's geometric convention is to place the most massive body on the left, and $\alpha$ is defined to be the counterclockwise angle between the binary axis and the source trajectory.  Initial model fits indicated significant deviations around the peak that are typically introduced when the angular radius of the source star is non-negligible relative to the angular size of the caustic.  We therefore investigated finite-source binary lens (FSBL) models, and took the limb-darkening of the source into account when computing the magnification of the source.  A linear limb-darkening model is commonly sufficient for microlensing models, and we adopt the widely-used formalism \citep{Albrow1999}: 

\begin{equation}
\label{eqn:limbdark1}
I_{\lambda} = \frac{F_{\lambda}}{\pi\theta_{*}^{2}} \left [ 1 - \Gamma_{\lambda} \left (1 - \frac{3}{2} \cos{\phi} \right ) \right ],
\end{equation}

where $I_{\lambda}$ is the intensity of the source at wavelength, $\lambda$, $F_{\lambda}$ is the total flux from the source in a given passband and $\phi$ is the angle between the line of sight to the observer and the normal to the stellar surface. The limb-darkening coefficient, $\Gamma_{\lambda}$ is related to the $u_{\lambda}$ limb-darkening coefficients derived from the ATLAS stellar atmosphere models presented \citep{Claret2011} by the expression:

\begin{equation}
\Gamma_{\lambda} = \frac{2 u_{\lambda}}{3 - u_{\lambda}}.
\end{equation}

The values of $u_{\lambda}$ and $\Gamma_{\lambda}$ applied for each dataset are presented in Table~\ref{tab:limbdark}.

\begin{table}[h!]
\centering
\caption{Linear limb-darkening coefficients used} \label{tab:limbdark}
\begin{tabular}{lllll}
\hline
\hline
Facility        & Filter         & $u_{\lambda}$   & $\Gamma_{\lambda}$ \\
\hline
LCO 1\,m        & SDSS-g$^{\prime}$ & 0.8852 & 0.8371 \\
LCO 1\,m        & SDSS-r$^{\prime}$ & 0.7311 & 0.6445 \\
LCO 1\,m        & SDSS-i$^{\prime}$ & 0.603  & 0.503 \\
Danish          & $i’_{\rm DK}$     & 0.5139 & 0.4134 \\
\hline
\hline
\end{tabular}
\end{table}

The PSF naturally differs between datasets acquired from different observing sites and instruments.  In the crowded star fields of the Galactic Bulge, the PSF of the source star is highly likely to be blended with those of neighboring stars.  The measured flux of the target at time $t$ in dataset $k$, $f(t,k)$ is calculated as a function of lensing magnification, $A(t)$, $f(t, k) = A(t)f_{s}(k) + f_{b}(k)$. Here, $f_{s}(k)$ is the flux of the source star and $f_{b}(k)$ represents the flux of all stars blended with the source in the data set. A regression fit was performed in the course of the modeling process to measure $f_{s}$ and $f_{b}$ for each data set. 

pyLIMA's Differential Evolution (DE, \citealt{StornPrice1997_DE}) solution-finding algorithm was used to explore parameter space and zero in on the region that best represents the data, after which we mapped the posterior distribution of each region using a Monte-Carlo Markov Chain (MCMC) algorithm (emcee; \citealt{ForemanMackey2013}). Once the parameter space minimum had been localized, the best-fitting model parameters were identified using the Levenberg-Marquardt algorithm \citep{Levenberg1944, Marquardt1963} or the Trust Region Reflective algorithm \citep{Colman1994,Branch1999}.  

\begin{figure}
\begin{tabular}{cc}
\includegraphics[width=0.5\linewidth]{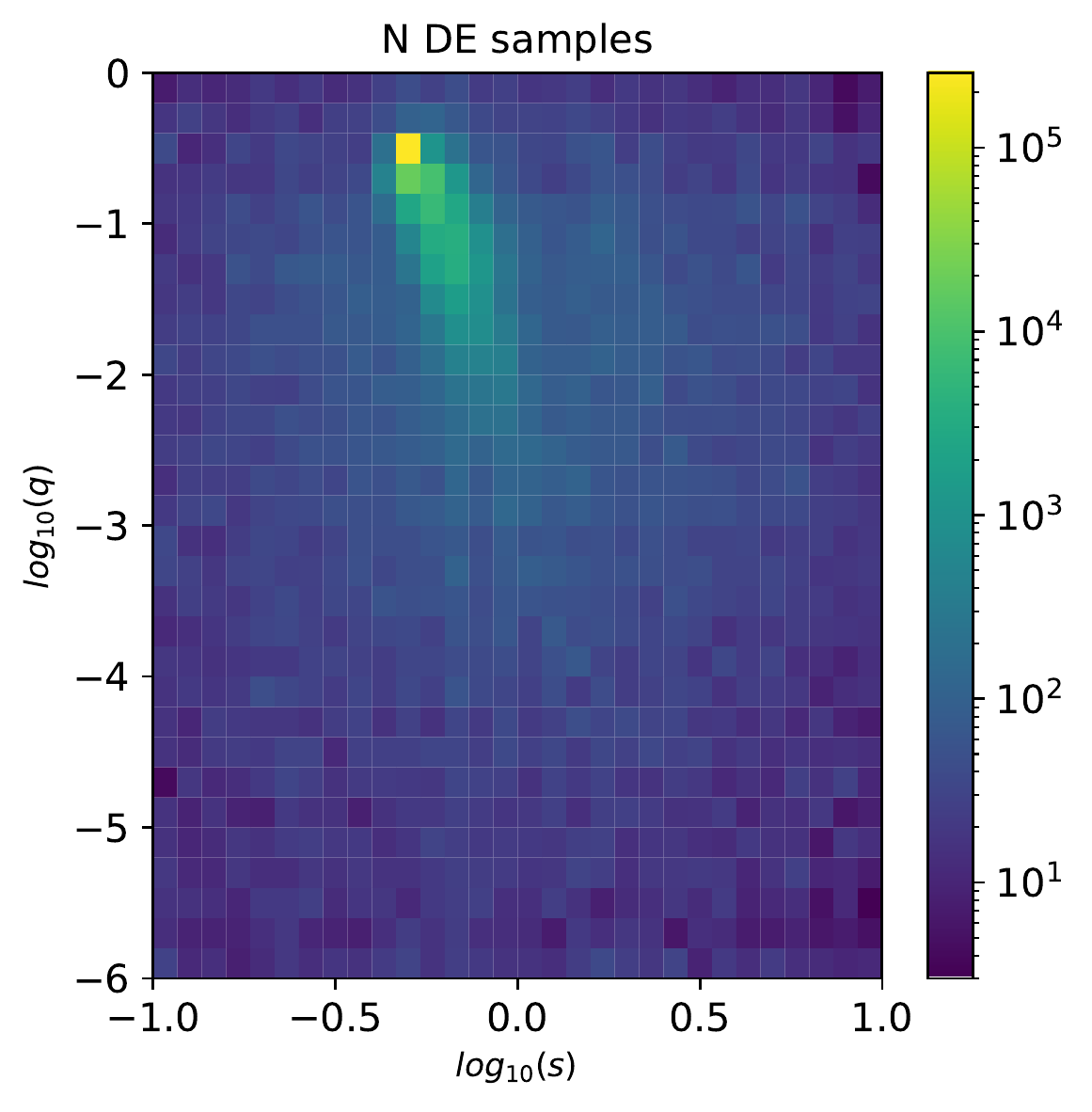}&
\includegraphics[width=0.5\linewidth]{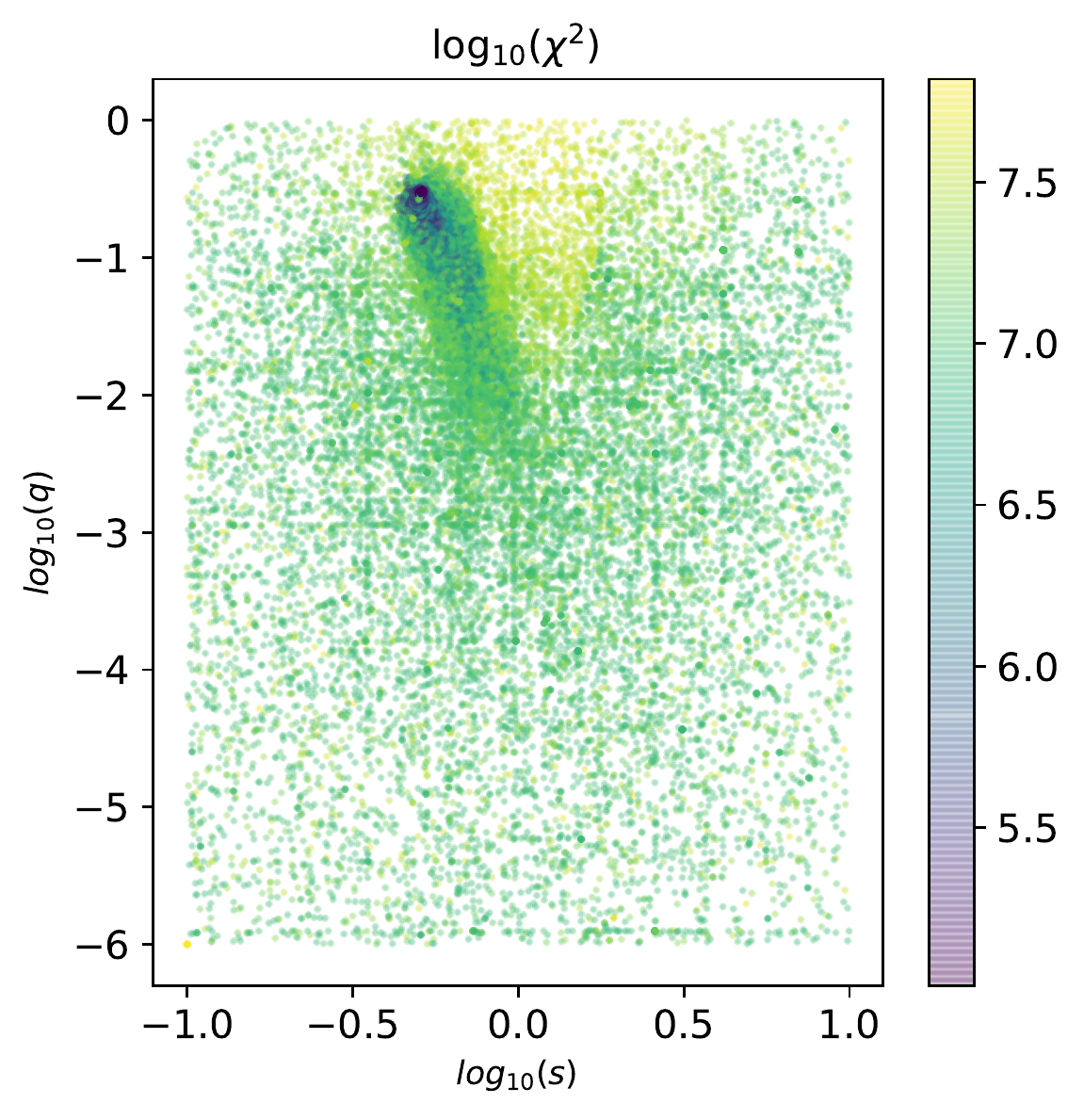}\\
\end{tabular}
\caption{Maps of the ($\log_{10}(s)$ vs. $\log_{10}(q)$) parameter space mapped out by a Differential Evolution (DE) algorithm.  (Left) A 2D histogram of the number of DE samples for each pixel in the parameter space, (right) a 2D histogram of the chi-square values of each pixel.  This plot was downsampled by a factor of 2 without loss of detail to minimize the plot filesize. \label{fig:de_maps}}
\end{figure}

The DE algorithm was initially given as few restrictions as possible ($t_{0}$ to lie within $\pm$50\,d of the event peak; -0.3 $\leq$ $u_{0}$ $\leq$ 0.3) so that it would explore a wide parameter space and identify all possible minima for further study.  The DE algorithm outputs the fit parameters for each candidate solution, which can be used to map the parameter space as shown in Figure~\ref{fig:de_maps}.  By design, the algorithm returns more solutions in regions where $\chi^{2}$ minima are located, so a 2D histogram of the number of solutions per element of $\log_{10}(s)$ vs. $\log_{10}(q)$ space indicates where solutions lie and further investigation is required.  This exploration indicated a single but extended minimum, and consistently converged on solutions where the source-lens relative trajectory intersected the central caustic.  The origin of the coordinate system was set to that of the central caustic during the modeling process, for increased stability of fit, so the impact parameter $u_{c}$, is measured relative to this point.

Before refining the model, we reviewed the photometric uncertainties for all light curves.  All photometry suffers from systematic noise at some level, and this must be quantified to avoid over-fitting the data.  pyLIMA provides statistical tests of the goodness-of-fit, including a Kolmogorov-Smirnov test, an Anderson-Darling test and a Shapiro-Wilk test \citep{Bachelet2015}.  If the $p$-value returned by tests was $<$1\%, the uncertainties on each dataset were revised, which was necessary in all cases. 

Following common practise (e.g. \citealt{Skowron2015}), we renormalized the photometric errors, $\sigma(k)$, of each dataset, $k$, according to the expression (in magnitude units):

\begin{equation}
\sigma(k)^{\prime} = \sqrt{a_{0}^{2}(k) + a_{1}^{2}(k) \sigma(k)^{2}}.
\end{equation}

The coefficients $a_{0}(k),a_{1}(k)$ were estimated by requiring that the reduced  $\chi^{2}_{\rm red} = 1$.  If the fit could not be constrained then the coefficients were set to 0.0 and 1.0, respectively. This could occur for a variety of reasons, the most common being that the majority of measurements in a given dataset were taken primarily over the peak of the event, where the rescaling fit was heavily influenced by residuals from the model, particularly around caustic crossings.  This was mitigated to some degree by iterating the model fitted with the rescaling process, to verify that the uncertainties of specific data points were not being excessively scaled.  A second problem was that the photometric uncertainties for a given dataset spanned a relatively short numerical range, leading to instability in the linear regression fit of the above function, and resulting in statistically nonsensical coefficients. Lastly, for some datasets the residual scatter in the photometry was accurately represented by the uncertainties, implying that no rescaling was required.  The adopted values are given in Table~\ref{tab:rescale_factors}.  

\begin{table}[h!]
\centering
\caption{Coefficients used in the rescaling of photometric uncertainties for each dataset and filter.} \label{tab:rescale_factors}
\begin{tabular}{lllllllll}
\hline
\hline
Facility                    & \multicolumn{2}{c}{\gp} & \multicolumn{2}{c}{\rp} & \multicolumn{2}{c}{\ip} & \multicolumn{2}{c}{$i’_{\rm DK}$} \\
                            & $a_{0}$ & $a_{1}$ & $a_{0}$ & $a_{1}$ & $a_{0}$ & $a_{1}$ & $a_{0}$ & $a_{1}$ \\
\hline
Chile, Dome A, fl15         & 0.0 & 1.0 & 0.0 & 1.0 & 0.0 & 1.0 & - & - \\ 
Chile, Dome C, fl03         & -   & -   & 0.0 & 1.0 & -   & -   & - & - \\ 
South Africa, Dome A, fl16  & 0.047$\pm$0.03 & 1.0 & 0.0 & 1.0 & 0.019$\pm$0.004 & 2.874$\pm$1.597 & - & - \\
South Africa, Dome C, fl06  & -   & -   & 0.0 & 1.0 & 0.0 & 1.0 & - & - \\ 
Australia, Dome A, fl12     & 0.0 &  1.0 & 0.0 &  1.0 & 0.0 &  2.445$\pm$0.831 & - & - \\ 
Australia, Dome B, fl11     & -   & -   & -   & -   & 0.0 & 1.0 & - & - \\ 
Danish, 1.54,m, DFOSC       & -   & -   & -   & -   & -   & -   & 0.0 &  1.0 \\
\hline
\hline
\end{tabular}
\end{table}

As there are both wide- and close-binary configurations that can produce very similar caustic structures (the well-documented close-wide degeneracy, \citealt{Dominik1999, Dominik2009}), we split the parameter space into two regions, $s>1$ and $s<1$, which were explored separately.  For this event, the close-binary solutions proved to be a significantly better fit than the wide-binary models; the parameters of the best fitting models in each case are presented in Tables~\ref{tab:closefitparams}--\ref{tab:widefitparams}.  We found a significant improvement in $\chi^{2}$ was achieved by including microlensing parallax, which is expected for an event of this duration, and also lens binary orbital motion.  

At each stage of modeling, as these effects were included, we explored finite-source, binary-lens (FSBL) models as well as USBL models.  While the best-fitting of these models indicated similar parameters to the USBL models, their $\chi^{2}$ values were found to be somewhat higher.  A close examination of the residuals showed that this is driven by a small number (5) of data points around the caustic crossing at 2458232.7, where the model is most sensitive to the limb-darkening of the source star. Two of the datapoints are in SDSS-g' band and 3 are in SDSS-i', which in principle might provide an independent constraint on $\Gamma$.  Regrettably, the caustic crossing occurred between the end of the night in Chile and the start of the night in Australia, and the data points were obtained from different instruments, under different conditions.  This is a situation where residual systematic noise in the photometry can easily exceed the finite source signature, so proceeding with finite-source models was judged to be unsafe.  

\begin{figure}
\epsscale{0.8}
\plotone{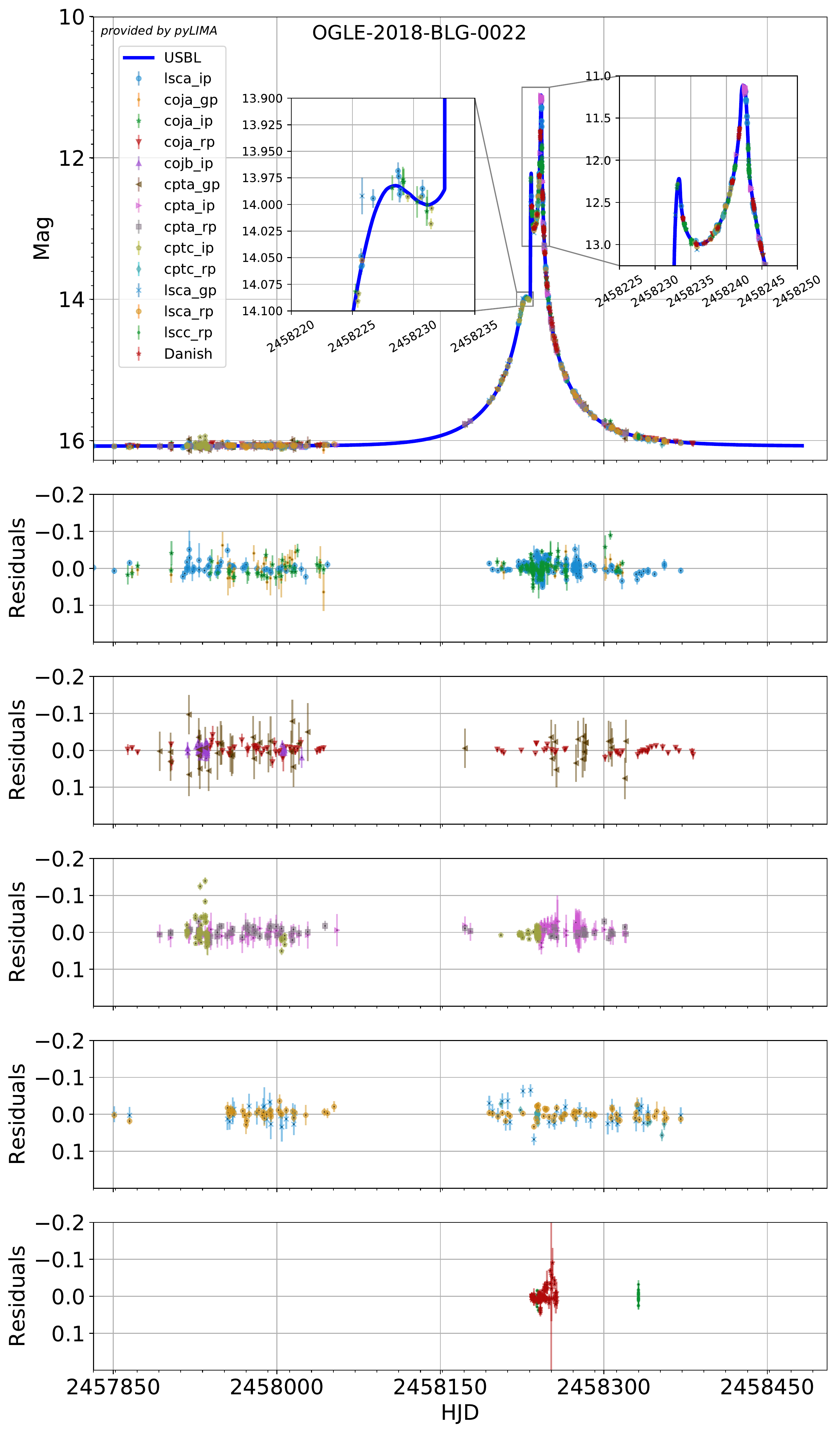}
\caption{Plot of all photometric datasets overlaid with the best-fitting USBL model light curve including parallax and orbital motion.  The inset in the top panel shows the light curve during the caustic crossing in more detail.  The bottom panel displays the photometric residuals once the model is subtracted from the data.  \label{fig:light curve}}
\end{figure}

\begin{figure}
\epsscale{0.8}
\plotone{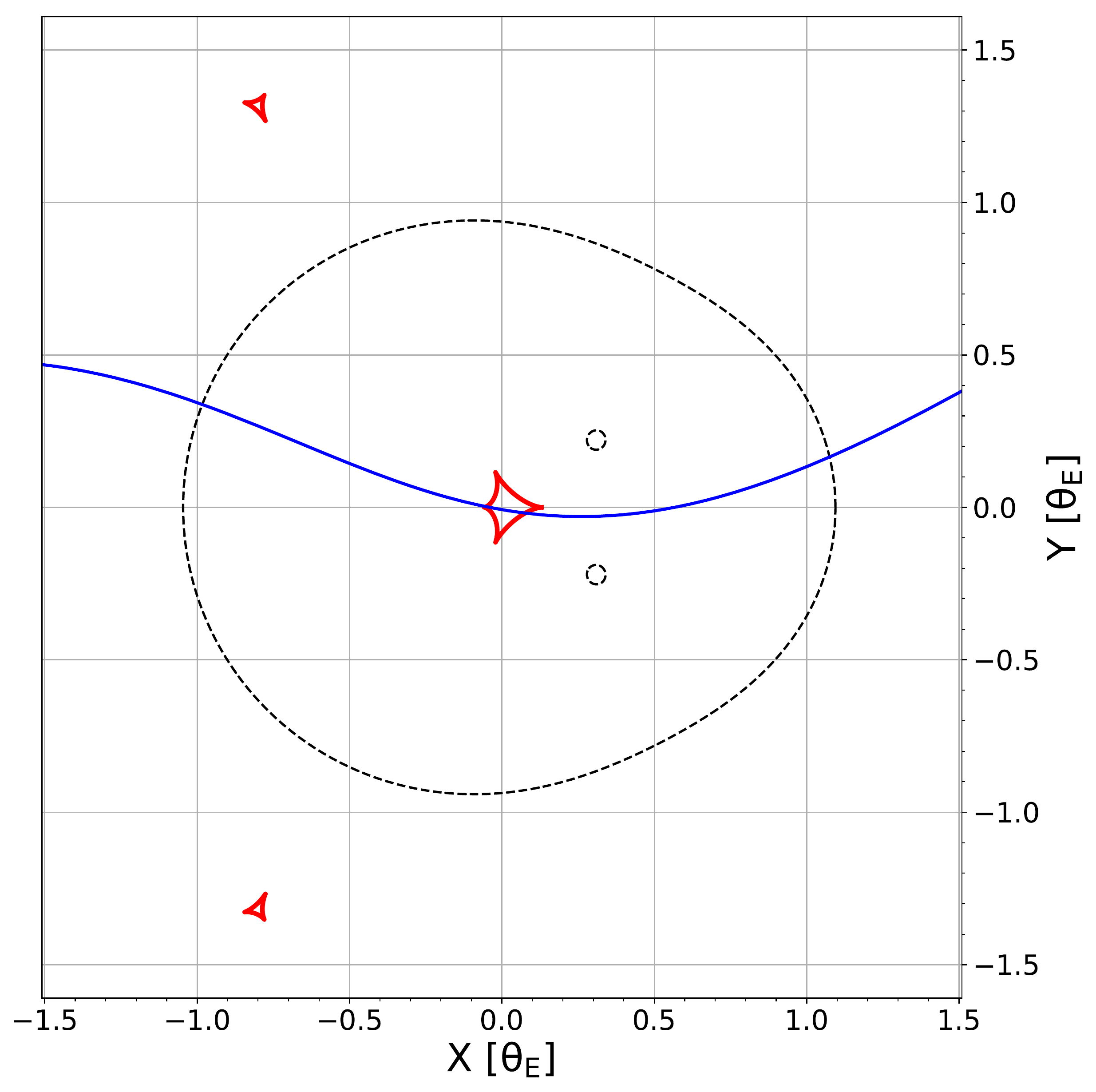}
\caption{Plot of the lens plane in a lens-centric geometry, showing the caustic structures for the binary lens in red in relation to the source-lens relative trajectory in blue.  The dotted black lines represent the critical curve.\label{fig:caustics}}
\end{figure}

Figure~\ref{fig:light curve} displays the light curve data overlaid with the best-fitting model, a uniform-source close-binary lens, and a plot of the source's trajectory relative to the lens plane and caustic structures is shown in Figure~\ref{fig:caustics}.  We note that there is a second degeneracy: lens-source relative trajectories with a negative $u_{c}$ value could in principle produce a very similar light curve.  These solutions were allowed during our fitting process, but were always disfavored in the results.  This would not strongly impact the physical characteristics of the lens inferred from the best-fit model. 

\begin{table}[h!]
\centering
\caption{Parameters of the best-fitting close-binary models.  Uncertainties are indicated in the second line for each parameter.  $\Delta \chi^{2}$ is calculated from the difference in $\chi^{2}$ between the each fit and the fit in its neighboring column to the left.} \label{tab:closefitparams}
\begin{tabular}{lllll}
\hline
\hline
Parameter           & Static binary & Binary+parallax  &\multicolumn{2}{c}{Binary+parallax}\\
                    &               &                  &\multicolumn{2}{c}{+orbital motion}\\
                    & USBL          & USBL           & USBL          & FSBL  \\
\hline
$t_{0}$ [HJD]        & 2458239.52808 & 2458239.98991 & 2458239.95128 & 2458240.02326  \\
                     & 0.00317       & 0.00326       & 0.00357       & 0.00381      \\
$u_{c}$              & 0.004421      & 0.004496      & 0.004426      & 0.004526       \\
                     & 0.000017      & 0.000015      & 0.000019      & 0.000017     \\
$t_{E}$ [days]       & 72.767        & 70.417        & 74.905        & 75.917        \\
                     & 0.065         & 0.069         & 0.051         & 0.071          \\
$\rho$               & 0.004207      & 0.004357      & 0.003967      & 0.004021      \\
                     & 0.000011      & 0.000013      & 0.0000089     & 0.000012     \\
$\log_{10}(s)$       & -0.29528      & -0.27221      & -0.27748      & -0.27484     \\
                     & 0.00017       & 0.00016       & 0.00023       & 0.00018        \\
$\log_{10}(q)$       & -0.48924      & -0.56329      & -0.58343      & -0.59851      \\
                     & 0.00030       & 0.00033       & 0.00024       & 0.00042      \\
$\alpha$ [radians]   & 2.97536       & 2.95992       & 2.96006       & 2.95522       \\
                     & 0.00032       & 0.00026       & 0.00034       & 0.00031        \\
$\pi_{\rm{E},N}$    &               & 0.5008        & 0.4718        & 0.4841        \\
                     &               & 0.0021        & 0.0015        & 0.0030         \\
$\pi_{\rm{E},E}$    &               & 0.0852        & 0.0664        & 0.069933      \\
                     &               & 0.0022        & 0.0018        & 0.0031        \\
$\frac{ds}{dt}$ [$\theta_{E}$/year]& &               & 0.001158      & 0.001552       \\
                     &               &               & 0.000039      & 0.000040      \\
$\frac{d\alpha}{dt}$ [radians/year]& &               & -0.000016     & -0.000442      \\
                     &               &               & 0.000038      & 0.000043      \\
$\chi^{2}$           & 16884.06      & 7239.03       & 6863.32       & 6979.39        \\
$\Delta \chi^{2}$    &               & -9645.03      & -375.71       & 116.07        \\
\hline
\end{tabular}
\end{table}

\begin{table}[h!]
\centering
\caption{Parameters of the best-fitting wide-binary models.  Uncertainties are indicated in the second line for each parameter. $\Delta \chi^{2}$ is calculated from the difference in $\chi^{2}$ between the each fit and the fit in its neighboring column to the left.} \label{tab:widefitparams}
\begin{tabular}{lllllll}
\hline
\hline
                     & Static binary    & Binary+parallax & Binary+parallax\\
                     &                  &               & +orbital motion\\
Parameter            & USBL             & USBL          & USBL          \\
\hline
$t_{0}$ [HJD]        & 2458238.66632    & 2458239.19101 & 2458240.07697 \\
                     & 0.00373          & 0.00352       & 0.04935       \\
$u_{c}$              & -0.002834        & -0.003548     & -0.006768     \\
                     & 0.000015         & 0.0000099     & 0.000027      \\
$t_{E}$ [days]       & 129.3778         & 101.455       & 122.516       \\
                     & 0.0072           & 0.091         & 0.107         \\
$\rho$               & 0.002042         & 0.002514      & 0.0025865     \\
                     & 0.000020         & 0.0000060     & 0.0000090     \\
$\log_{10}(s)$       & 0.578942         & 0.51043       & 0.44685       \\
                     & 0.000026         & 0.00013       & 0.00037       \\
$\log_{10}(q)$       & -0.00065         & -0.13962      & -0.33554      \\
                     & 0.00036          & 0.00063       & 0.00098       \\
$\alpha$ [radians]   & -3.01041         & -2.99934      & -1.89562      \\
                     & 0.00016          & 0.00023       & 0.00132       \\
$\pi_{\rm{E},N}$    &                  & -0.28192      & -0.48863      \\
                     &                  & 0.00077       & 0.00315       \\
$\pi_{\rm{E},E}$    &                  & 0.1125        & -0.3915       \\
                     &                  & 0.0011        & 0.0018        \\
$\frac{ds}{dt}$ [$\theta_{E}$/year]&    &               & -0.022095     \\
                     &                  &               & 0.000044       \\
$\frac{d\alpha}{dt}$ [radians/year]&    &               & -0.007852     \\
                     &                  &               & 0.000013      \\
$\chi^{2}$           & 36501.669        & 20519.11      & 8141.82       \\
$\Delta \chi^{2}$    &                  & -15982.56     & -12377.29     \\
\hline
\end{tabular}
\end{table}

\section{Source Color Analysis}
\label{sec:colour_analysis}

We adopted data from Chile, Dome A, camera fl15 to act as our photometric reference, since this site consistently has the best observing conditions of the whole network.  After reviewing all available data, a trio of single \gp, \rp, \ip\  images taken sequentially on 2017 July 26 between 04:05 -- 04:17 UTC were selected as the reference images for these datasets because they were obtained in the best seeing, transparency and sky background conditions.  These images were used as the reference images for the DIA pipeline.  PSF fitting photometry was conducted on the same images, in order to determine the reference fluxes of all detected stars.  

\begin{figure}
\plottwo{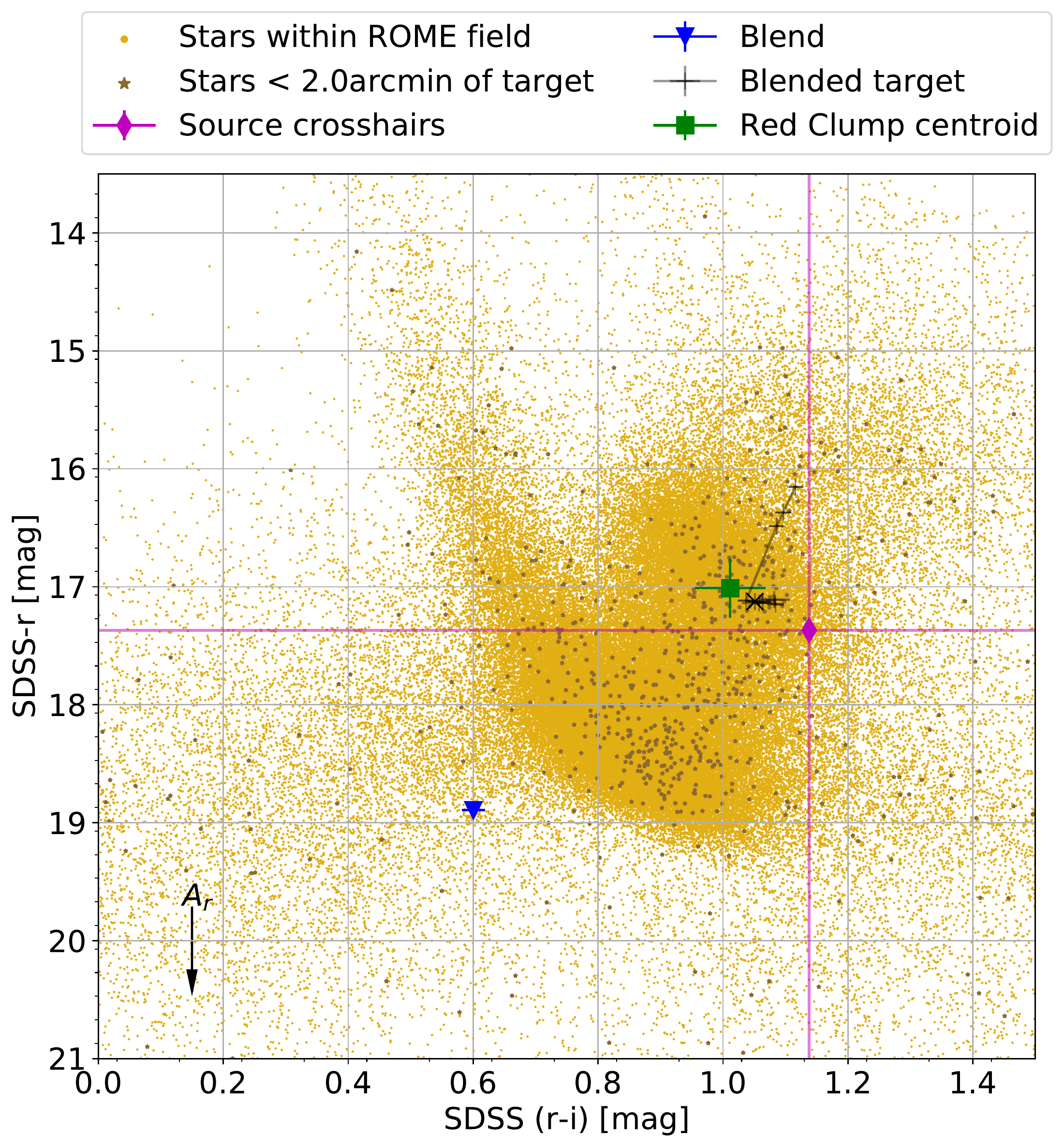}{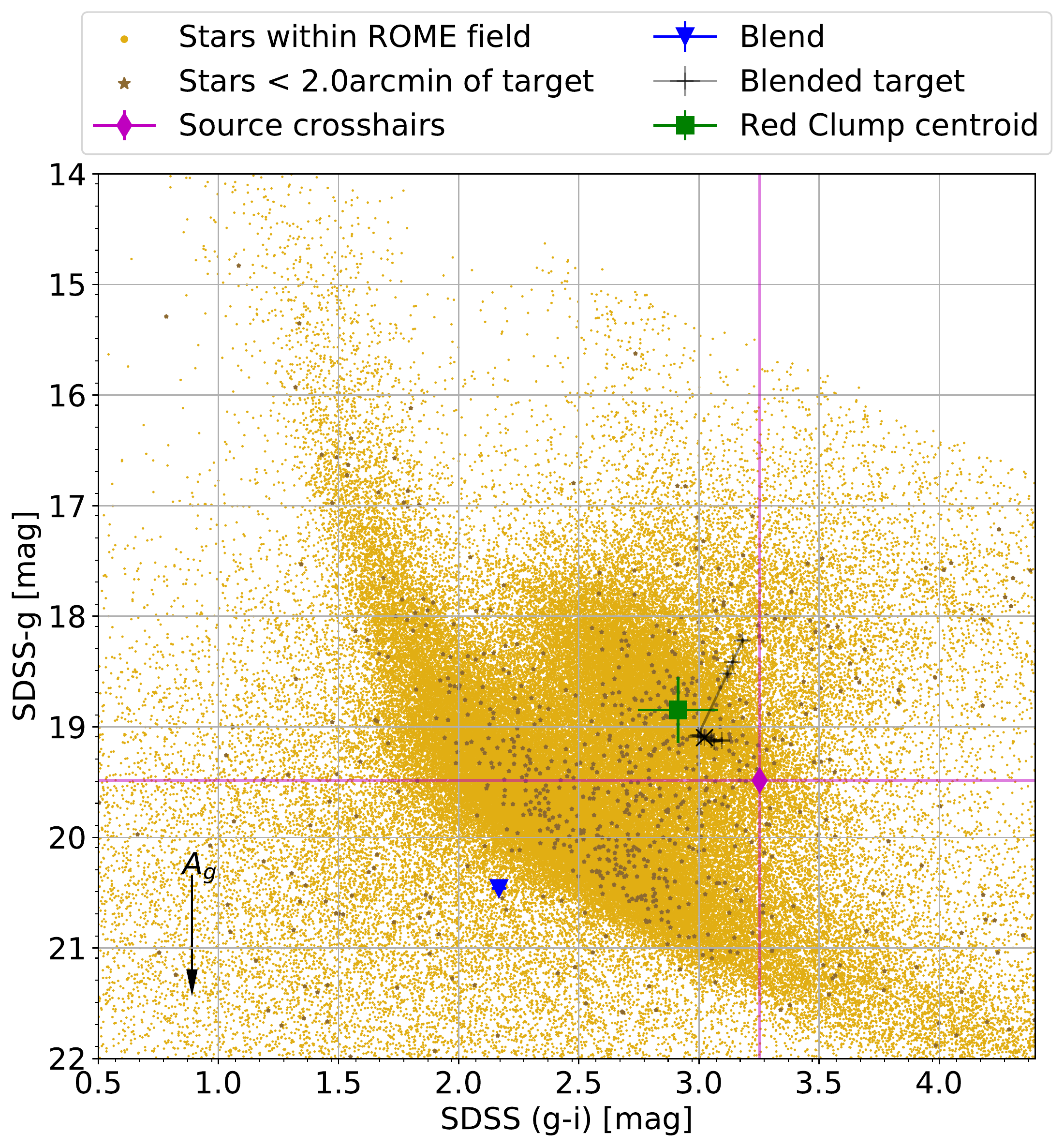}
\caption{ROME color-magnitude diagrams for the field containing OGLE-2018-BLG-0022.  Stars within 2\,arcmin of the target are highlighted in dark brown, whereas stars in the rest of the field are plotted in yellow.  The center of the Red Clump is marked with a green square, and that of the blend by a blue triangle.  The magnitude of the source+blend is plotted as a function of time as black $+$ symbols.  The magenta diamond marks the location of the source, overlaid with crosshairs.  An arrow indicates the extinction vector at the distance of the lens.  \label{fig:cmds}}
\end{figure}

The positions of all detected stars (as determined from the World Coordinate System fit, WCS, for each image) were cross-matched against the VPHAS+ catalog \citep{Drew2014VPHAS+}, from which calibrated SDSS-g, -r and -i magnitudes were extracted.  To mitigate the impact of differential extinction across the field of view, stars within 2\,arcmin of the lensed star were selected for the purpose of measuring the photometric transformation from LCO instrumental magnitudes to the VPHAS+ system.  Color-magnitude diagrams from the ROME data are presented in Figure~\ref{fig:cmds}.  

While this procedure provides an approximate photometric calibration, fields in the Galactic Bulge suffer from high extinction, which is often spatially variable across the field of view of a single ROME exposure.  To account for this, it has become standard practice in microlensing to measure the offset of the Red Clump from its expected magnitude and color.  

Red Clump giant stars are often used as standard candles, since their absolute luminosity is constant, being relatively insensitive to changes in metalicity and age, and they occur with high frequency across the Galactic Plane.  Recently, \cite{Ruiz-Dern2018} summarized Red Clump photometric properties in a wide range of photometric systems, including their absolute magnitudes in the SDSS passbands: $M_{g,RC,0} = 1.331\pm0.056$\,mag, $M_{r,RC,0}$ = 0.552$\pm$0.026\,mag, $M_{i,RC,0} = 0.262\pm0.032$\,mag.  

To determine the apparent magnitude and colors of the Red Clump stars, we assume that they are located in the Galactic Bar.  \cite{Nataf2013} indicated that the Galactic Bar is orientated at a viewing angle of $\phi_{\rm{Bar}}$ = 40$^{\circ}$, meaning that the distance to the Red Clump, $D_{RC}$, is a function of Galactic longitude, $l$:

\begin{equation}
\frac{R_{0}}{D_{RC}} = \frac{\sin{\phi_{\rm{Bar}}} + l}{\phi_{\rm{Bar}}} = \cos{l} + \sin{l}\cot{\phi_{\rm{Bar}}},
\end{equation}

\noindent where $R_{0}$ = 8.16\,kpc (we note that the bar angle may be somewhat smaller, \citealt{Cao2013, Wegg2013}, but this will cause only small changes to the results).  Based on the location of OGLE-2018-BLG-0022 in Galactic coordinates $(l,b)$ = (1.82295, -2.44338)$^{\circ}$, the distance to the Red Clump in this field is estimated to be 7.87\,kpc.  We used this to estimate the apparent photometric properties (denoted by $m_{\lambda,RC,0}$ for different passbands, $\lambda$) summarized in Table~\ref{tab:RCproperties}.  

The Red Clump is clearly identifiable in the ROME color-magnitude diagrams (Fig.~\ref{fig:cmds}).  Stars within 2\,arcmin of the target were used to measure the centroid of the Clump in magnitude and color by applying the following selection cuts: $15.5 \leq i \leq 16.5$\,mag, $16.2 \leq r \leq 17.5$\,mag, $17.8 \leq g \leq 19.5$\,mag, $0.8 \leq (r-i) \leq 1.2$\,mag, $1.5 \leq (g-r) \leq 2.2$\,mag.  The measured centroids of the Red Clump are presented in Table~\ref{tab:RCproperties}.  

\begin{table}[h!]
\centering
\caption{Photometric properties of the Red Clump, with absolute magnitudes  ($M_{\lambda}$)} taken from \cite{Ruiz-Dern2018}, and the measured properties from ROME data. \label{tab:RCproperties}
\begin{tabular}{ll}
\hline
\hline
$M_{g,RC,0}$ & 1.331 $\pm$ 0.056\,mag\\
$M_{r,RC,0}$ & 0.552 $\pm$ 0.026\,mag\\
$M_{i,RC,0}$ & 0.262 $\pm$ 0.032\,mag\\
$(g-r)_{RC,0}$ & 0.779 $\pm$ 0.062\,mag\\
$(g-i)_{RC,0}$ & 1.069 $\pm$ 0.064\,mag\\
$(r-i)_{RC,0}$ & 0.290 $\pm$ 0.041\,mag\\
$m_{g,RC,0}$ & 15.810 $\pm$ 0.056\,mag\\
$m_{r,RC,0}$ & 15.031 $\pm$ 0.026\,mag\\
$m_{i,RC,0}$ & 14.741 $\pm$ 0.032\,mag\\
$m_{g,RC,\rm{centroid}}$  & 18.85 $\pm$ 0.30\,mag\\
$m_{r,RC,\rm{centroid}}$  & 17.01 $\pm$ 0.25\,mag\\
$m_{i,RC,\rm{centroid}}$  & 16.03 $\pm$ 0.24\,mag\\
$(g-r)_{RC,\rm{centroid}}$ & 1.87  $\pm$ 0.13\,mag\\
$(r-i)_{RC,\rm{centroid}}$ & 1.01 $\pm$ 0.06\,mag\\
$A_{g}$ & 3.037 $\pm$ 0.056\,mag\\
$A_{r}$ & 1.981 $\pm$ 0.026\,mag\\
$A_{i}$ & 1.290 $\pm$ 0.032\,mag\\
$E(g-r)$ & 1.091 $\pm$ 0.062\,mag\\
$E(r-i)$ & 0.722 $\pm$ 0.041\,mag\\
\hline
\end{tabular}
\end{table}

The offset of the Red Clump from its expected photometric properties was used to estimate the extinction, $A_{\lambda}$, and reddening, $E(\rm{color})$ for the Red Clump along the line of sight to the target. 

These quantities were then used to correct the photometric properties of the source and blend, as derived from the best-fitting light curve model, assuming that they have the same extinction and reddening as the Red Clump.  The resulting data are summarized in Table~\ref{tab:targetphot}. 

\begin{table}[h!]
\centering
\caption{Photometric properties of the source star (S) and blend (b).} \label{tab:targetphot}
\begin{tabular}{llll}
\hline
\hline
$m_{g,\rm S}$ & 19.484 $\pm$ 0.007\,mag & $m_{g,b}$ & 20.462 $\pm$ 0.027\,mag\\
$m_{r,\rm S}$ & 17.369 $\pm$ 0.002\,mag & $m_{r,b}$ & 18.895 $\pm$ 0.013\,mag\\
$m_{i,\rm S}$ & 16.231 $\pm$ 0.001\,mag & $m_{i,b}$ & 18.294 $\pm$ 0.013\,mag\\
$(g-r)_{\rm S}$ & 2.115 $\pm$ 0.007\,mag & $(g-r)_{b}$ & 1.567 $\pm$ 0.030\,mag\\
$(g-i)_{\rm S}$ & 3.253 $\pm$ 0.007\,mag & $(g-i)_{b}$ & 2.168 $\pm$ 0.030\,mag\\
$(r-i)_{\rm S}$ & 1.138 $\pm$ 0.002\,mag & $(r-i)_{b}$ & 0.601 $\pm$ 0.018\,mag\\
$m_{g,\rm S,0}$ & 16.447 $\pm$ 0.056\,mag &  & \\
$m_{r,\rm S,0}$ & 15.388 $\pm$ 0.026\,mag &  & \\
$m_{i,\rm S,0}$ & 14.941 $\pm$ 0.032\,mag &  & \\
$(g-r)_{\rm S,0}$ & 1.059 $\pm$ 0.007\,mag &  & \\
$(g-i)_{\rm S,0}$ & 1.506 $\pm$ 0.007\,mag &  & \\
$(r-i)_{\rm S,0}$ & 0.447 $\pm$ 0.002\,mag &  & \\
\hline
\end{tabular}
\end{table}

We note that the ROME survey strategy provides a useful means to verify the source flux determined from the model.  Since ROME observations are always conducted as a sequence of back-to-back ($g^{\prime}$,$r^{\prime}$,$i^{\prime}$) exposures taken within $\sim$15\,mins of each other, the magnification of the event can normally be taken to be approximately the same for all 3 images in a trio (excluding caustic crossings).  These observations can be used to measure the source color and blend flux independently of the model, as follows.  The total flux measured in a given passband $\lambda$, $f_{\lambda}$ consists of the source flux, $f_{\rm S,\lambda}$, multiplied by the lensing magnification, $A$, combined with the flux from any other blended stars along the line of sight, $f_{b,\lambda}$: $f_{\lambda}(t) = f_{\rm S,\lambda} A(t) + f_{b,\lambda}$. Contemporaneous fluxes in multiple passbands can be combined as:

\begin{equation}
f_{\lambda,1}(t) = \frac{f_{\rm S,\lambda,1}}{f_{\rm S,\lambda,2}}(f_{\lambda,2}(t) - f_{b,\lambda,2}) + f_{b,\lambda,1}.
\end{equation}

This allows the source color to be measured by linear regression from the slope of the fluxes in different passbands, plotted against one another.  Applying this technique, we measured: $(g-r)_{\rm S}$ = 2.115$\pm$0.007\,mag, $(g-i)_{\rm S}$ = 3.253$\pm$0.007\,mag, $(r-i)_{\rm S}$ = 1.138$\pm$0.002\,mag.  These values are consistent with the colors determined from the model-predicted source fluxes in Table~\ref{tab:targetphot} The resulting timeseries of source color measurements are show in Fig.~\ref{fig:cmds}, and can be evaluated relative to the crosshairs indicating the source color measured from the light curve analysis.  

\begin{figure}
\plotone{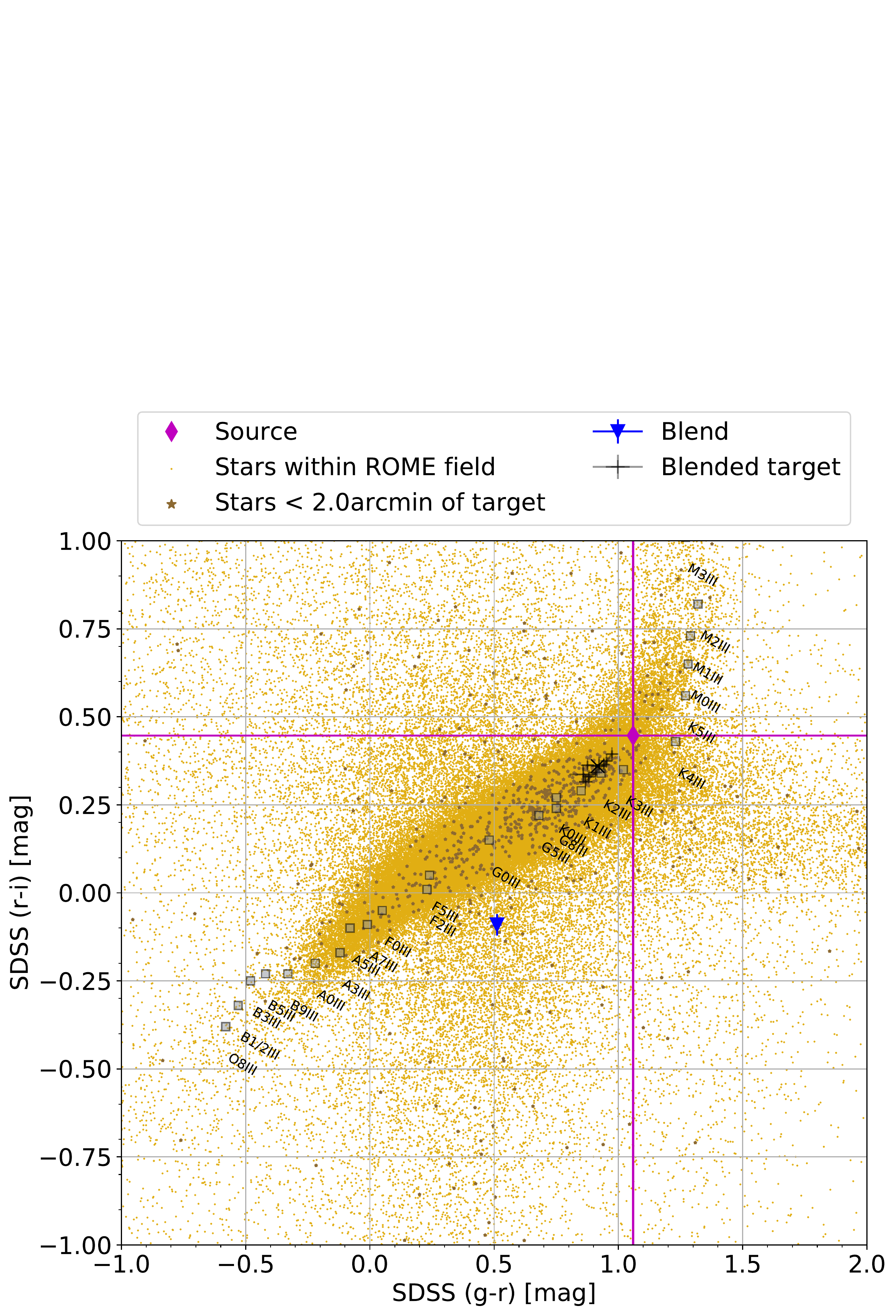}
\caption{ROME color-color diagram for the field containing OGLE-2018-BLG-0022.  Stars within 2\,arcmin of the target are highlighted in dark brown, whereas stars in the rest of the field are plotted in yellow.  The location of the source is indicated by a magenta diamond, and that of the blend with a blue triangle.  The magnitude of the source+blend is plotted as a function of time as black $+$ symbols.  The overlaid colors of the giant sequence was derived from \cite{Pickles1998}, and plotted for comparison. \label{fig:colcold}}
\end{figure}

The source star's location on the color-color diagram (Figure~\ref{fig:colcold}) was compared with theoretical stellar isochrones derived from the {\sc parsec} model\footnote{http://stev.oapd.inaf.it/cgi-bin/cmd} \citep{Bressan2012} for solar metallicity and ages ranging from $3.98 \times 10^{6}$ to $1.26 \times 10^{10}$\,yrs, to find the closest matching colors for each isochrone.  This analysis indicated a source effective temperature of $T_{\rm eff} = 4290.9 \pm 50.0$\,K, suggesting that the source star is a K-type star.  

The angular radius, $\theta_{S}$, for the source star was then calculated using the relationships between the limb-darkened $\theta_{S}$ and stellar colors in SDSS passbands derived from \cite{Boyajian2014}. Both color indices for which coefficients were published yielded consistent estimates: $\theta_{\rm S,(g-r)} = 7.144 \pm 0.319\,\mu$as and $\theta_{S,(g-i)} = 7.431 \pm 0.232\,\mu$as.  We adopt an average of these two results, $\theta_{S} = 7.288 \pm 0.394\,\mu$as.

The final required constraint is to determine the distance to the source, so that $\rho = \theta_{S}/\theta_{E}$ can be used to measure the Einstein radius.  The best way to provide this constraint would be a measurement of the source's parallax from the {\em Gaia} mission \citep{GaiaMission}, although its source catalog is restricted to the brightest stars only in  Bulge fields, owing to limitations of the on-board processing.  The {\em Gaia} Data Release 2 catalog \citep{Gaia_DR2} reported a source (id=4062576103277425536) within 0.185\,arcsec of this event, though its parallax measurement (0.13233847489290135$\pm$0.10027593536929187\,mas) was flagged as uncertain.  The catalog of distances provided by \cite{Bailer-Jones2018} gave an ill-constrained measurement of $8342_{-5333}^{+12860}$\,pc.  It should also be borne in mind that this measurement reflects the flux of the source+blend, at baseline, and the methodology was not optimized for crowded fields.  

However, the source angular radii derived from \cite{Boyajian2014} and the color indices imply that the source is a giant, and its position on the color-magnitude diagrams is consistent with a Red Clump giant in the Bulge at a distance of 7.87\,kpc, as calculated earlier.  Adopting this distance for the source, we infer a radius of 12.330$\pm$0.666\,$R_{\odot}$.  

Combining these quantities, with the parameters of our best-fitting model, we infer the physical properties of the lens from the following relations.  The angular Einstein radius, $\theta_{E}$, was extracted from the ratio of source radii $\rho$ in Einstein and $\theta_{\rm S}$ in absolute units.  This quantity relates directly to the total lens mass, $M_{L}$, the lens distance, $D_{L}$ and the lens-source separation, $D_{LS}$.   

\begin{eqnarray}
\theta_{\rm{E}} &=& \frac{\theta_{\rm S}}{\rho} = \sqrt{ \frac{4GM_{\rm L}}{\mu D_{\rm L}c^{2}} },\\
\mu &=& 1 + \frac{D_{\rm L}}{D_{\rm{LS}}},
\end{eqnarray}

The distance to the lens was inferred from the relative parallax, $\pi_{rel}$, determined from our best-fit model, and the parallax to the source, $\pi_{S}$:

\begin{eqnarray}
\pi_{\rm L} &=& \pi_{\rm{rel}} + \pi_{\rm S},\\
\pi_{\rm S} &=& 1 \textrm{AU} / D_{\rm S},\\
D_{\rm L} &=& 1 \textrm{AU} / \pi_{\rm L}.
\end{eqnarray}

The resulting lens properties are summarized in Table~\ref{tab:lensproperties}.  The lens masses are consistent with a low-mass stellar binary composed of an M6-7 star orbiting an M3 star.  

\begin{table}[h!]
\centering
\caption{Physical properties of the source and lens system.} \label{tab:lensproperties}
\begin{tabular}{lll}
\hline
\hline
Parameter   &   Units    &   Value \\
$\theta_{\rm{S}}$  & $\mu$as     & 7.288$\pm$0.394\\
$\theta_{\rm{E}}$  & $\mu$as     & 1837.145$\pm$99.384\\
$R_{\rm{S}}$       & $R_{\odot}$ & 12.327$\pm$0.666\\
$M_{L,tot}$        & $M_{\odot}$ & 0.473$\pm$0.026\\
$M_{L,1}$          & $M_{\odot}$ & 0.376$\pm$0.020\\
$M_{L,2}$          & $M_{\odot}$ & 0.098$\pm$0.005\\
$D_{L}$            & Kpc         & 0.998$\pm$0.047\\
$a_{\perp}$       & AU          & 0.967$\pm$0.070\\
$\mu$              & mas yr$^{-1}$ & 8.96$\pm$0.48\\
\hline
\end{tabular}
\end{table}

\section{Assessment of the lens and blended flux}
\label{sec:lens}

The lensing system in this case is relatively close, compared with other microlensing discoveries, and its location suggests that the binary may lie in the Galactic Disk.  Given the measured masses, the simplest explanation is that the lens consists of two main sequence components.  However, we noted that, with a distance modulus of 9.99$\pm$0.47\,mag, a main sequence binary might be detectable, and we estimated its likely photometric properties as follows.  

We extracted the absolute magnitudes of M-type stars from a PARSEC isochrone, assuming solar age and metallicity, and calculated the expected apparent magnitudes of the binary at the lens distance (see Table~\ref{tab:binaryphot}).  These magnitudes are significantly brighter than the limiting magnitude of the ROME data (limited by the sky background, $\sim$21.969\,mag [SDSS-g], $\sim$21.989\,mag [SDSS-r], $\sim$22.010\,mag [SDSS-i]), and suggest that the lens could be contributing to the blend flux we measured from the light curve.  

\begin{figure}
\plotone{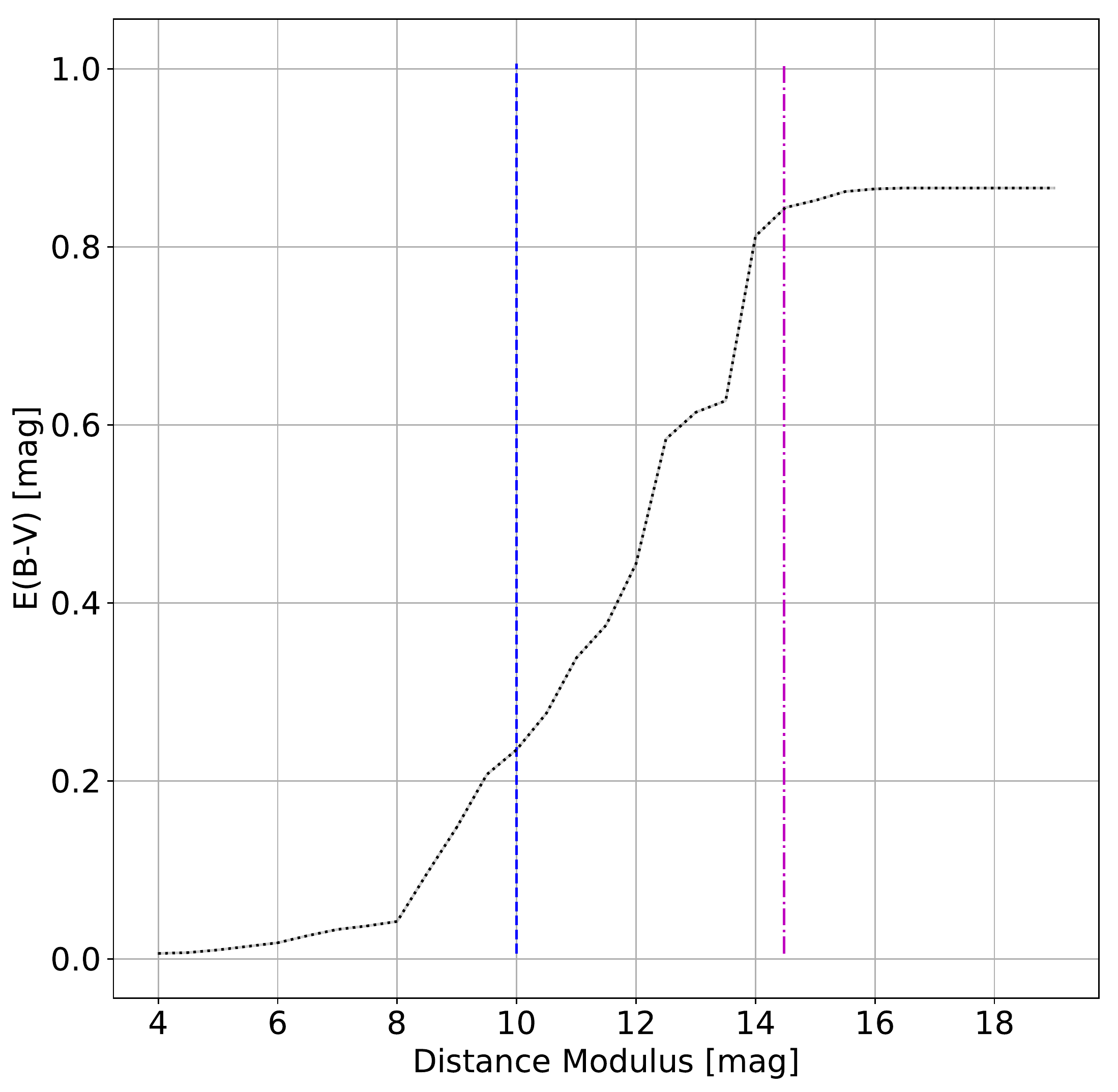}
\caption{Colour excess as a function of distance modulus along the line of sight to OGLE-2018-BLG-0022, derived from the 3D extinction maps published by \cite{Green2015}.  The purple dash-dotted line marks the distance modulus of the source star while the blue dashed line indicates that of the lens. \label{fig:3Dreddening}}
\end{figure}

Before drawing any conclusions however, extinction and reddening must be considered. Data from the Pan-STARRS1 \citep{PanSTARRS1} and 2MASS \citep{2MASS} surveys have been combined to provided maps of the 3-dimensional reddening within the Milky Way \citep{Green2015}, which we can use to estimate this quantity along the line of sight to the source star in this event.  By interpolating the data at the $(l,b)$ of this event, we estimated the colour excess to the lens star to be $E(B-V) = 0.235\pm0.032$\,mag (Fig.~\ref{fig:3Dreddening}).  This was used to estimate the extinction in $V$-band, $A_{V} = R_{V} E(B-V)$, where the reddening, $R_{V}$ was estimated for the Galactic Bulge by \cite{Nataf2013} to be $\sim$2.5$\pm$0.2.  We therefore found $A_{V}$ = 0.59$\pm$0.09\,mag.  This was used to estimate the extinction in Sloan filters by applying the transforms derived by \cite{SchlaflyFinkbeiner2011}, interpolating between the discrete values of $R_{V}$ they provided to arrive at extinction values for this field of $A_{(SDSS-g)}$ = 0.851$\pm$0.118\,mag, $A_{(SDSS-r)}$ = 0.532$\pm$ 0.074\,mag, $A_{(SDSS-i)}$ = 0.382$\pm$0.053\,mag.  The extinction-corrected apparent magnitudes for the lens binary scenario are presented at the bottom of Table~\ref{tab:binaryphot}. 

\begin{table}[h!]
\centering
\caption{Predicted photometric properties of the lens system.  Apparent magnitudes are calculated for the measured lens distance without extinction or reddening, except for the bottom section.  } \label{tab:binaryphot}
\begin{tabular}{lccc}
\hline
\hline
Quantity	  & M3-dwarf   & M7-dwarf   & MS-binary \\
$[mag]$	      &		       & 	        & M3$+$M7     \\
$M_{B}$       & 13.175 & 21.124 & 13.174\\
$M_{V}$       & 11.574 & 18.674 & 11.572\\
$M_{g}$       & 11.933 & 19.149 & 11.932\\
$M_{r}$       & 10.409 & 17.181 & 10.407\\
$M_{i}$       & 9.475 & 14.700 & 9.466\\
$M_{J}$       & 7.566 & 11.033 & 7.522\\
$M_{H}$       & 7.014 & 10.458 & 6.969\\
$M_{Ks}$      & 6.779 & 10.178 & 6.733\\
$(B-V)$       & 1.601 & 2.450 & 1.602\\
$(g-r)$       & 1.524 & 1.968 & 1.525\\
$(r-i)$       & 0.934 & 2.481 & 0.941\\
$(J-H)$       & 0.552 & 0.575 & 0.553\\
$(H-Ks)$      & 0.235 & 0.280 & 0.237\\
$(J-Ks)$      & 0.787 & 0.855 & 0.790\\
$m_{B}$       & 23.169 & 31.118 & 23.168\\
$m_{V}$       & 21.568 & 28.668 & 21.566\\
$m_{g}$       & 21.927 & 29.143 & 21.926\\
$m_{r}$       & 20.403 & 27.175 & 20.401\\
$m_{i}$       & 19.469 & 24.694 & 19.460\\
$m_{J}$       & 17.560 & 21.027 & 17.516\\
$m_{H}$       & 17.008 & 20.452 & 16.963\\
$m_{Ks}$      & 16.773 & 20.172 & 16.727\\
\hline
$m_{V,corr}$  & 22.155 & 29.255 & 22.153\\
$(B-V)_{corr}$& 1.836 & 2.685 & 1.837\\
$m_{g,corr}$  & 22.778 & 29.994 & 22.777\\
$m_{r,corr}$  & 20.935 & 27.707 & 20.933\\
$m_{i,corr}$  & 19.851 & 25.076 & 19.842\\
$(g-r)_{corr}$& 1.843 & 2.287 & 1.844\\
$(r-i)_{corr}$& 1.084 & 2.631 & 1.091\\
\hline
\end{tabular}
\end{table}

The measured blend photometry in Table~\ref{tab:targetphot} indicates one or more objects that are significantly brighter than the photometry predicted for a main sequence M3+M7 binary, implying that the light originates from a separate object(s) -- a common situation in the crowded star fields of the Galactic Bulge.  Nevertheless, we note that the lens should be easily detectable in 2--4\,m-class telescopes, particularly in the NIR.  

\section{Conclusions}
The microlensing event OGLE-2018-BLG-0022 revealed the presence of an M3+M7 binary star, previously undetected owing to its intrinsically low luminosity.  That said, the binary in this event is unusually close to the Earth for a microlens -- $\sim$1\,kpc away -- and the object shows a correspondingly high relative proper motion of 8.96\,mas yr$^{-1}$.  This makes it a good candidate for high spatial resolution AO imaging in the relatively near future which, as discussed by \cite{Henderson2014}, could provide an independent verification of the lens mass determination.  While the proximity of the lens, resulting in a large (1.84\,mas) angular Einstein radius, would have been resolvable to interferometry, as demonstrated by \cite{Dong2018}, the source star in this case was too faint for current instruments.  The discovery highlight's microlensing's capability to map populations beyond the solar neighborhood that would otherwise be hidden by their intrinsically faint luminosities.  




\section{Acknowledgements}
RAS and EB gratefully acknowledge support from NASA grant NNX15AC97G.  YT and JW acknowledge the support of DFG priority program SPP 1992 “Exploring the Diversity of Extrasolar Planets” (WA 1047/11-1).  KH acknowledges support from STFC grant ST/R000824/1.  This research has made use of NASA's Astrophysics Data System, and the NASA Exoplanet Archive.  The work was partly based on data products from observations made with ESO Telescopes at the La Silla Paranal Observatory under programme ID 177.D-3023, as part of the VST Photometric H{alpha} Survey of the Southern Galactic Plane and Bulge (VPHAS+, www.vphas.eu).  This work also made use of data from the European Space Agency (ESA) mission {\it Gaia} (\url{https://www.cosmos.esa.int/gaia}), processed by the {\it Gaia} Data Processing and Analysis Consortium (DPAC, \url{https://www.cosmos.esa.int/web/gaia/dpac/consortium}). Funding for the DPAC
has been provided by national institutions, in particular the institutions
participating in the {\it Gaia} Multilateral Agreement.  CITEUC is funded by National Funds through FCT - Foundation for Science and Technology (project: UID/Multi/00611/2013) and FEDER - European Regional Development Fund through COMPETE 2020 – Operational Programme Competitiveness and Internationalization (project: POCI-01-0145-FEDER-006922).  The work by SS and SR was supported by a grant (95843339) from the Iran National Science Foundation (INSF).  DMB acknowledges the support of the NYU Abu Dhabi Research Enhancement Fund under grant RE124.  This research uses data obtained through the Telescope Access Program (TAP), which has been funded by the National Astronomical Observatories of China, the Chinese Academy of Sciences, and the Special Fund for Astronomy from the Ministry of Finance. This work was partly supported by the National Science Foundation of China (Grant No. 11333003, 11390372 and 11761131004 to SM).

\bibliographystyle{plainnat}  % needs package natbib
\bibliography{ob180022_rome}

%\begin{thebibliography}{}

%\end{thebibliography}



\end{document}